\documentclass{article}
\usepackage{ijcai18}

\usepackage{times}
\usepackage{xcolor}
\usepackage{soul}
\usepackage[utf8]{inputenc}
\usepackage[small]{caption}
\usepackage{hyperref}
\usepackage{enumitem}
\usepackage{graphicx}
\usepackage{amsmath}
\usepackage{amsfonts}
\usepackage{subfigure}
\hypersetup{draft} 
\usepackage{titlesec}
\titlespacing{\paragraph}{0pt}{0.5\baselineskip}{1em}

\title{Outer Product-based Neural Collaborative Filtering}

\author{
Xiangnan He$^1$, 
Xiaoyu Du$^{1,2}$, 
Xiang Wang$^1$, 
Feng Tian$^3$,
Jinhui Tang$^4$ {\normalfont and} Tat-Seng Chua$^1$
\\ 
$^1$ National University of Singapore \\
$^2$ Chengdu University of Information Technology\\
$^3$ Northeast Petroleum University\\
$^4$ Nanjing University of Science and Technology\\
\{xiangnanhe, duxy.me\}@gmail.com, xiangwang@u.nus.edu, dcscts@nus.edu.sg
}
\begin{document}

\maketitle

\begin{abstract}
	In this work, we contribute a new multi-layer neural network architecture named ONCF to perform collaborative filtering. The idea is to use an outer product to explicitly model the pairwise correlations between the dimensions of the embedding space. In contrast to existing neural recommender models that combine user embedding and item embedding via a simple concatenation or element-wise product, our proposal of using outer product above the embedding layer results in a two-dimensional \textit{interaction map} that is more expressive and semantically plausible. Above the interaction map obtained by outer product, we propose to employ a convolutional neural network to learn high-order correlations among embedding dimensions. Extensive experiments on two public implicit feedback data demonstrate the effectiveness of our proposed ONCF framework, in particular, the positive effect of using outer product to model the correlations between embedding dimensions in the low level of multi-layer neural recommender model. \footnote{Work appeared in IJCAI 2018. The experiment codes are available at: \href{https://github.com/duxy-me/ConvNCF}{https://github.com/duxy-me/ConvNCF}}
\end{abstract}

\section{Introduction}
\label{sec:introduction}

To facilitate the information seeking process for users in the age of data deluge, various information retrieval (IR) technologies have been widely deployed~\cite{garcia2011information}. As a typical paradigm of information push, recommender systems have become a core service and a major monetization method for many customer-oriented systems~\cite{wang2018path}. 
Collaborative filtering (CF) is a key technique to build a personalized recommender system, which infers a user's preference not only from her behavior data but also the behavior data of other users. Among the various CF methods, 
model-based CF, more specifically, matrix factorization based methods~\cite{rendle2009bpr,he2016fast,DCF} are known to provide superior performance over others and have become the mainstream of recommendation research. 

The key to design a CF model is in 1) how to represent a user and an item, and 2) how to model their interaction based on the representation. As a dominant model in CF, matrix factorization (MF) represents a user (or an item) as a vector of latent factors (also termed as \textit{embedding}), and models an interaction as the inner product between the user embedding and item embedding. Many extensions have been developed for MF from both the modeling perspective~\cite{Suhang2015,Yu:2018:ACR,TEM} and learning perspective~\cite{rendle2009bpr,iCD,he2018adversarial}. For example, DeepMF~\cite{DeepMF} extends MF by learning embeddings with deep neural networks,
BPR~\cite{rendle2009bpr} learns MF from implicit feedback with a pair-wise ranking objective, and the recently proposed adversarial personalized ranking (APR)~\cite{he2018adversarial} employs an adversarial training procedure to learn MF. 

Despite its effectiveness and many subsequent developments, we point out that MF has an inherent limitation in its model design. Specifically, it uses a fixed and data-independent function --- i.e., the inner product --- as the interaction function~\cite{he2017neural}.
As a result, it essentially assumes that the embedding dimensions (i.e., dimensions of the embedding space) are independent with each other and contribute equally for the prediction of all data points. 
This assumption is impractical, since the embedding dimensions could be interpreted as certain properties of items~\cite{zhang2014explicit}, which are not necessarily to be independent. 
Moreover, this assumption has shown to be sub-optimal for learning from real-world feedback data that has rich yet complicated patterns, since several recent efforts on neural recommender models~\cite{tay2018latent,bai2017neural} have demonstrated that better recommendation performance can be obtained by learning the interaction function from data. 

Among the neural network models for CF, neural matrix factorization (NeuMF)~\cite{he2017neural} provides state-of-the-art performance by complementing the inner product with an adaptable multiple-layer perceptron (MLP) in learning the interaction function. Later on, using multiple nonlinear layers above the embedding layer has become a prevalent choice to learn the interaction function. Specifically, two common designs are placing a MLP above the concatenation~\cite{he2017neural,bai2017neural} and the element-wise product~\cite{zhang2017joint,wang2017item} of user embedding and item embedding. We argue that a potential limitation of such two designs is that there are few correlations between embedding dimensions being modeled. Although the following MLP is theoretically capable of learning any continuous function according to the universal approximation theorem~\cite{hornik1991approximation}, there is no practical guarantee that the dimension correlations can be effectively captured with current optimization techniques. 

In this work, we propose a new architecture for neural collaborative filtering~(NCF) by integrating the correlations between embedding dimensions into modeling. Specifically, we propose to use an outer product operation above the embedding layer, explicitly capturing the pairwise correlations between embedding dimensions. We term the correlation matrix obtained by outer product as the \textit{interaction map}, which is a $K\times K$ matrix where $K$ denotes the embedding size. The interaction map is rather suitable for the CF task, since it not only subsumes the interaction signal used in MF (its diagonal elements correspond to the intermediate results of inner product), but also includes all other pairwise correlations. Such rich semantics in the interaction map facilitate the following non-linear layers to learn possible high-order dimension correlations. Moreover, the matrix form of the interaction map makes it feasible to learn the interaction function with the effective convolutional neural network (CNN), which is known to generalize better and is more easily to go deep than the fully connected MLP. 

The contributions of this paper are as follows. 
\begin{itemize}[leftmargin=*]
	\item We propose a new neural network framework ONCF, which supercharges NCF modeling with an outer product operation to model pairwise correlations between embedding dimensions. 
	\item We propose a novel model named ConvNCF under the ONCF framework, which leverages CNN to learn high-order correlations among embedding dimensions from locally to globally in a hierarchical way. 
	\item We conduct extensive experiments on two public implicit feedback data, which demonstrate the effectiveness and rationality of ONCF methods. 
	\item This is the first work that uses CNN to learn the interaction function between user embedding and item embedding. It opens new doors of exploring the advanced and fastly evovling CNN methods for recommendation research.  
\end{itemize}

\section{Proposed Methods}
We first present the \textbf{O}uter product based \textbf{N}eural \textbf{C}ollaborative \textbf{F}iltering~(ONCF) framework.
We then elaborate our proposed \textbf{Conv}olutional \textbf{NCF}~(ConvNCF) model, an instantiation of ONCF that uses CNN to learn the interaction function based on the interaction map. Before delving into the technical details, we first introduce some basic notations. 

Throughout the paper, we use bold uppercase letter (e.g., $\textbf{P}$) to denote a matrix, bold lowercase letter to denote a vector (e.g., $\textbf{p}_u$), and calligraphic uppercase letter to denote a tensor (e.g., $\mathcal{S}$).
Moreover, scalar $p_{u,k}$ denotes the $(u,k)$-th element of matrix $\textbf{P}$, and vector $\textbf{p}_u$ denotes the $u$-th row vector in $\textbf{P}$.
Let $\mathcal{S}$ be 3D tensor, then scalar $s_{a,b,c}$ denotes the $(a,b,c)$-th element of tensor $\mathcal{S}$, and vector $\textbf{s}_{a,b}$ denotes the slice of $\mathcal{S}$ at the element $(a,b)$.

\subsection{ONCF framework}
\begin{figure}
	\includegraphics[width=\linewidth]{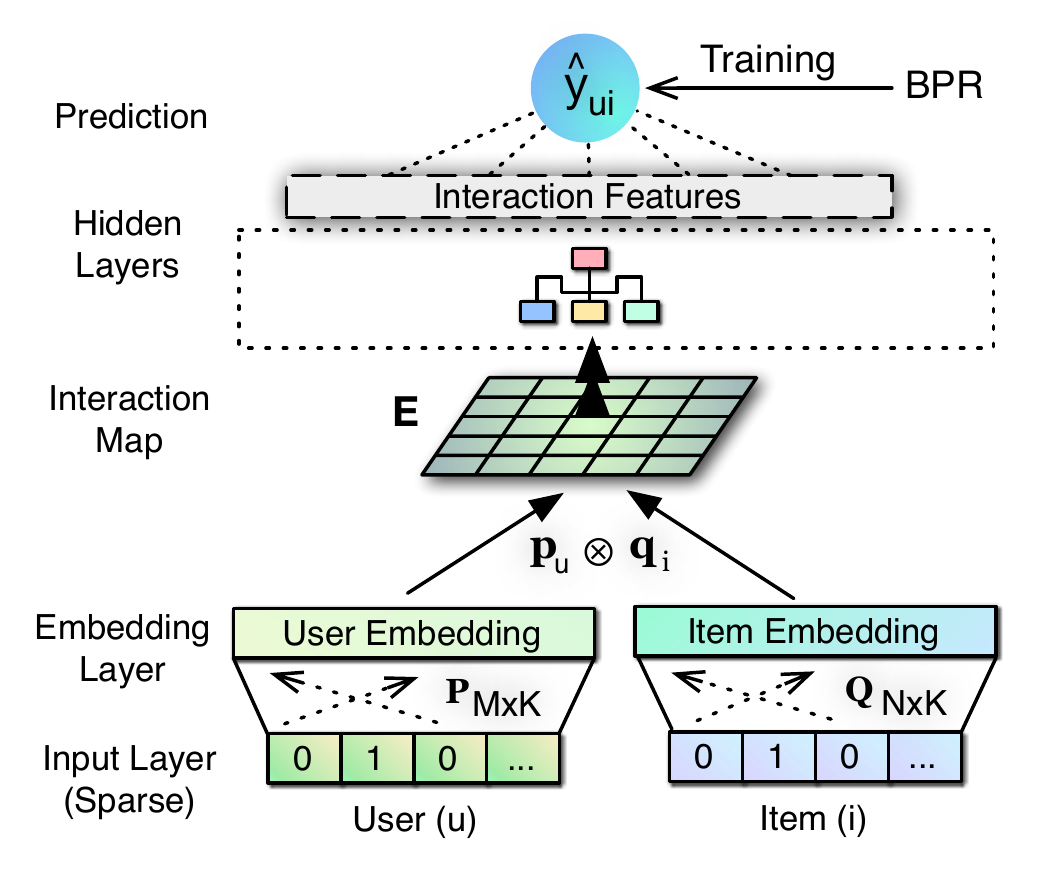} \vspace{-20pt}
	\caption{Outer Product-based NCF framework}\vspace{-10pt}
	\label{fig:framework}
\end{figure}

Figure~\ref{fig:framework} illustrates the ONCF framework.
The target of modeling is to estimate the matching score between user $u$ and item $i$, i.e., $\hat{y}_{ui}$; and then we can generate a personalized recommendation list of items for a user based on the scores. 
\vspace{+5pt} 

\paragraph{Input and Embedding Layer.} Given a user $u$ and an item $i$ and their features (e.g., ID, user gender, item category etc.), we first employ one-hot encoding on their features. 
Let $\textbf{v}_u^U$ and $\textbf{v}_i^I$ be the feature vector for user $u$ and item $i$, respectively, we can obtain their embeddings $\textbf{p}_u$ and $\textbf{q}_i$ via
\begin{equation}\small
    \textbf{p}_u = \textbf{P}^T \textbf{v}_u^U, \quad  \textbf{q}_i = \textbf{Q}^T \textbf{v}_i^I,
\end{equation}
where $\textbf{P}\in\mathbb{R}^{M\times K}$ and $\textbf{Q}\in\mathbb{R}^{N\times K}$ are the embedding matrix for user features and item features, respectively; $K, M,$ and $N$ denote the embedding size, number of user features, and number of item features, respectively.  
Note that in the pure CF case, only the ID feature will be used to describe a user and an item~\cite{he2017neural}, and thus $M$ and $N$ are the number of users and number of items, respectively. \vspace{+5pt}

\begin{figure*}[t]
    \center
	\includegraphics[width=0.9\linewidth]{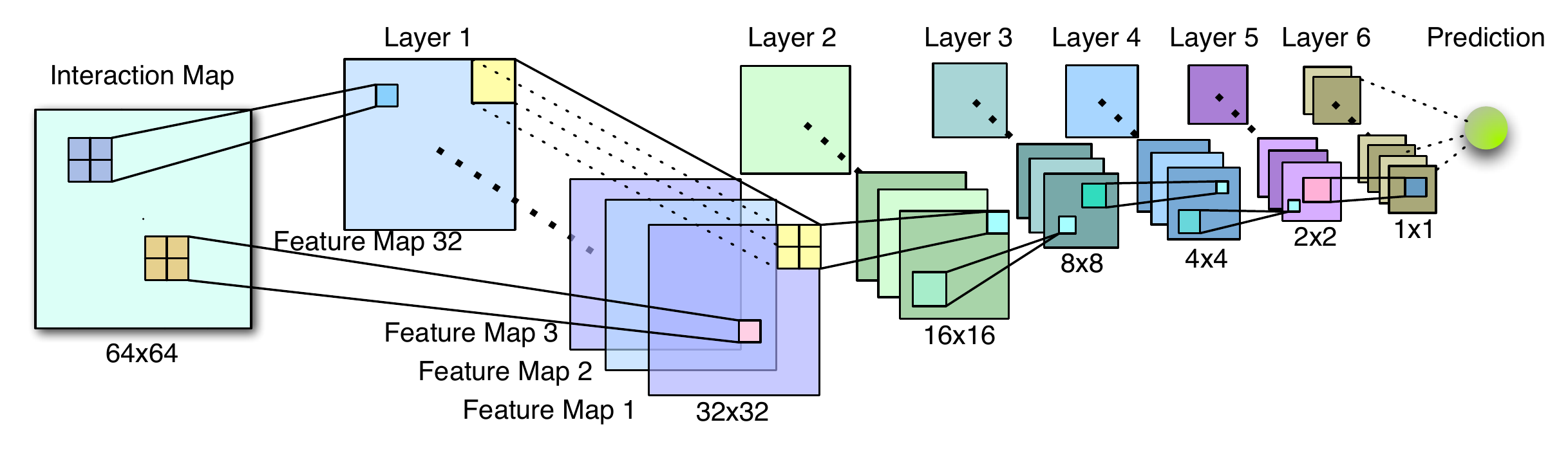} \vspace{-10pt}
	\caption{An example of the architecture of our ConvNCF model that has 6 convolution layers with embedding size 64.}\vspace{-10pt}
	\label{fig:ConNCF}
\end{figure*}

\paragraph{Interaction Map.} Above the embedding layer, we propose to use an outer product operation on $\textbf{p}_u$ and $\textbf{q}_i$ to obtain the interaction map:
\begin{equation}\small
    \textbf{E} = \textbf{p}_u \otimes \textbf{q}_i = \textbf{p}_u \textbf{q}_i^T,
\end{equation}
where \textbf{E} is a $K \times K$ matrix, in which each element is evaluated as: $e_{k_1, k_2} = p_{u, k_1} q_{i, k_2}$. 

This is the core design of our ONCF framework to ensure the effectiveness of ONCF for the recommendation task. Compared to existing recommender systems~\cite{he2017neural,zhang2017joint}, we argue that using outer product is more advantageous in threefold: 1) it subsumes matrix factorization (MF) --- the dominant method for CF --- which considers only diagonal elements in our interaction map; 2) it encodes more signal than MF by accounting for the correlations between different embedding dimensions; and 3) it is more meaningful than the simple concatenation operation, which only retains the original information in embeddings without modeling any correlation. 
Moreover, it has been recently shown that, modeling the interaction of feature embeddings explicitly is particularly useful for a deep learning model to generalize well on sparse data, whereas using concatenation is sub-optimal~\cite{NFM,LatentCross}. 

Lastly, another potential benefit of the interaction map lies in its 2D matrix format --- which is the same as an image. In this respect, the pairwise correlations encoded in the interaction map can be seen as the local features of an ``image''. As we all know, deep learning methods achieve the most success in computer vision domain, and many powerful deep models especially the ones based on CNN (e.g., 
ResNet~\cite{he2016deep} and DenseNet~\cite{huang2017densely}) have been developed for learning from 2D image data. Building a 2D interaction map allows these powerful CNN models to be also applied to learn the interaction function for the recommendation task. \vspace{+5pt}

\paragraph{Hidden Layers.} Above the interaction map is a stack of hidden layers, which targets at extracting useful signal from the interaction map. It is subjected to design and can be abstracted as $\textbf{g} = f_\Theta(\textbf{E})$, where $f_\Theta$ denotes the model of hidden layers that has parameters $\Theta$, and $\textbf{g}$ is the output  vector to be used for the final prediction. 
Technically speaking, $f_\Theta$ can be designed as any function that takes a matrix as input and outputs a vector. In Section~\ref{ssec:cncf}, we elaborate how CNN can be employed to extract signal from the interaction map. \vspace{+5pt}

\paragraph{Prediction Layer.} The prediction layer takes in vector $\textbf{g}$ and outputs the prediction score as: $\hat{y}_{ui} = \textbf{w}^T \textbf{g}$, where vector $\textbf{w}$ re-weights the interaction signal in $\textbf{g}$. To summarize, the model parameters of our ONCF framework are $\Delta = \{\textbf{P}, \textbf{Q}, \Theta, \textbf{w} \}$.

\subsubsection{Learning ONCF for Personalized Ranking}
Recommendation is a personalized ranking task. To this end, we consider learning parameters of ONCF with a ranking-aware objective. In the NCF paper~\cite{he2017neural}, the authors advocate the use of a pointwise classification loss to learn models from implicit feedback. 
However, another more reasonable assumption is that observed interactions should be ranked higher than the unobserved ones. To implement this idea, \cite{rendle2009bpr} proposed a Bayesian Personalized Ranking (BPR) objective function as follows:
\begin{equation}\small\label{eq:bpr}
    L(\Delta) = \sum_{(u,i,j)\in \mathcal{D}} -\ln \sigma(\hat{y}_{ui} - \hat{y}_{uj}) + \lambda_{\Delta} ||\Delta||^2, 
\end{equation}
where $\lambda_{\Delta}$ are parameter specific regularization hyperparameters to prevent overfitting, and $\mathcal{D}$ denotes the set of training instances: $\mathcal{D}:= \{(u,i,j)| i\in \mathcal{Y}_u^+ \wedge j\notin  \mathcal{Y}_u^+ \}$, where $\mathcal{Y}_u^+$ denotes the set of items that has been consumed by user $u$.
By minimizing the BPR loss, we tailor the ONCF framework for correctly predicting the relative orders between interactions,
rather than their absolute scores as optimized in pointwise loss~\cite{he2017neural,he2016fast}. This can be more beneficial for addressing the personalized ranking task. 

It is worth pointing out that in our ONCF framework, the weight vector $\textbf{w}$ can control the magnitude of the value of $\hat{y}_{ui}$ for all predictions. As a result, scaling up $\textbf{w}$ can increase the margin $\hat{y}_{ui} - \hat{y}_{uj}$ for all training instances and thus decrease the training loss. To avoid such trivial solution in optimizing ONCF, it is crucial to enforce $L_2$ regularization or the max-norm constraint on $\textbf{w}$. Moreover, we are aware of other pairwise objectives have also been widely used for personalized ranking, such as the L2 square loss~\cite{wang2017item}.
We leave this exploration for ONCF as future work, as our initial experiments show that optimizing ONCF with the BPR objective leads to good top-$k$ recommendation performance. 

\subsection{Convolutional NCF}
\label{ssec:cncf}

\paragraph{Motivation: Drawback of MLP.} In ONCF, the choice of hidden layers has a large impact on its performance. A straightforward solution is to use the MLP network as proposed in NCF~\cite{he2017neural}; note that to apply MLP on the 2D interaction matrix $\textbf{E}\in \mathbb{R}^{K\times K}$, we can flat $\textbf{E}$ to a vector of size $K^2$.
Despite that MLP is theoretically guaranteed to have a strong representation ability~\cite{hornik1991approximation}, its main drawback of having a large number of parameters can not be ignored. 
As an example, assuming we set the embedding size of a ONCF model as 64 (i.e., $K=64$) and follow the common practice of the half-size tower structure.
In this case, even a 1-layer MLP has $8,388,608$ (i.e., $4,096\times 2,048$) parameters, not to mention the use of more layers. We argue that such a large number of parameters makes MLP prohibitive to be used in ONCF because of three reasons: 1) It requires powerful machines with large memories to store the model; and 2) It needs a large number of training data to learn the model well; and 3) It needs to be carefully tuned on the regularization of each layer
to ensure good generalization performance\footnote{In fact, another empirical evidence is that most papers used MLP with at most 3 hidden layers, and the performance only improves slightly (or even degrades) with more layers~\cite{he2017neural,Covington:2016,NFM}}. \vspace{+5pt}

\paragraph{The ConvNCF Model.} To address the drawback of MLP, we propose to employ CNN above the interaction map to extract signals. As CNN stacks layers in a locally connected manner, it utilizes much fewer parameters than MLP. This allows us to build deeper models than MLP easily, and benefits the learning of high-order correlations among embedding dimensions. Figure \ref{fig:ConNCF} shows an illustrative example of our ConvNCF model. Note that due to the complicated concepts behind CNN (e.g., stride, padding etc.), we are not ambitious to give a systematic formulation of our ConvNCF model here. Instead, without loss of generality, we explain ConvNCF of this specific setting, since it has empirically shown good performance in our experiments. Technically speaking, any structure of CNN and parameter setting can be employed in our ConvNCF model. 
First, in Figure \ref{fig:ConNCF}, the size of input interaction map is $64\times 64$, and the model has 6 hidden layers, where each hidden layer has 32 feature maps. 
A feature map $c$ in hidden layer $l$ is represented as a 2D matrix $\textbf{E}^{lc}$; since we set the stride to 2, the size of $\textbf{E}^{lc}$ is half of its previous layer $l-1$, e.g. $\textbf{E}^{1c}\in \mathbb{R}^{32\times 32}$ and $\textbf{E}^{2c}\in \mathbb{R}^{16\times 16}$. All feature maps of Layer $l$ can be represented as a 3D tensor $\mathcal{E}^{l}$.

Given the input interaction map $\textbf{E}$, we can first get the feature maps of Layer 1 as follows:
\begin{equation}\small
\begin{aligned}
\mathcal{E}^{1} &= [e^{1}_{i,j,c}]_{32\times 32\times 32}, \quad \text{where} \\
e^{1}_{i,j,c} &= \text{ReLU}(b_1 + \sum_{a=0}^1 \sum_{b=0}^1
 e_{2i+a,2j+b}\cdot \underbrace{  t^{1}_{1-a,1-b,c}}_{\text{convolution filter}}),
\end{aligned}
\end{equation}
where $b_1$ denotes the bias term for Layer 1, and $\mathcal{T}^{1}=[t^{1}_{a,b,c}]_{2
\times 2\times 32}$ is a 3D tensor denoting the convolution filter for generating feature maps of Layer 1. We use the rectifer unit as activation function, a common choice in CNN to build deep models. 
Following the similar convolution operation, we can get the feature maps for the following layers. The only difference is that from Layer 1 on, the input to the next layer $l+1$ becomes a 3D tensor $\mathcal{E}^l$:
\begin{equation}\small
\begin{aligned}
\mathcal{E}^{l+1} &= [e^{l+1}_{i,j,c}]_{s\times s\times 32}, \quad \text{where}\  1\leq l\leq 5, \ s = \frac{64}{2^{l+1}}, \\ e^{l+1}_{i,j,c} &= \text{ReLU}(b_{l+1} + \sum_{a=0}^1 \sum_{b=0}^1
 \textbf{e}^l_{2i+a,2j+b}\cdot \textbf{t}^{l+1}_{1-a,1-b,c}),
\end{aligned}
\end{equation}
where $b_{l+1}$ denotes the bias term for Layer $l+1$, and  $\mathcal{T}^{l+1}=[t^{l+1}_{a,b,c,d}]_{2\times 2\times 32\times 32}$ denote the 4D convolution filter for Layer $l+1$. The output of the last layer is a tensor of dimension $1\times 1 \times 32$, which can be seen as a vector and is projected to the final prediction score with a weight vector $\textbf{w}$. 

Note that convolution filter can be seen as the ``locally connected weight matrix'' for a layer, since it is shared in generating all entries of the feature maps of the layer. This significantly reduces the number of parameters of a convolutional layer compared to that of a fully connected layer. Specifically, in contrast to the 1-layer MLP that has over 8 millions parameters, the above 6-layer CNN has only about 20 thousands parameters, which are several magnitudes smaller. This makes our ConvNCF more stable and generalizable than MLP. 

\vspace{+5pt}

\paragraph{Rationality of ConvNCF.} Here we give some intuitions on how ConvNCF can capture high-order correlations among embedding dimensions. In the interaction map $\textbf{E}$, each entry $e_{ij}$ encodes the second-order correlation between the dimension $i$ and $j$. Next, each hidden layer $l$ captures the correlations of a $2\times 2$ local area\footnote{The size of the local area is determined by our setting of the filter size, which is subjected to change with different settings.} of its previous layer $l-1$. As an example, the entry $e^{1}_{x,y,c}$ in Layer 1 is dependent on four elements $[e_{2x,2y}; e_{2x,2y+1}; e_{2x+1,2y}; e_{2x+1, 2y+1}]$, which means that it captures the 4-order correlations among the embedding dimensions $[2x; 2x+1; 2y; 2y+1]$. Following the same reasoning process, each entry in hidden layer $l$ can be seen as capturing the correlations in a local area of size $2^{l}$ in the interaction map $\textbf{E}$. 
As such, an entry in the last hidden layer encodes the correlations among all dimensions. Through this way of stacking multiple convolutional layers, we allow ConvNCF to learn high-order correlations among embedding dimensions from locally to globally, based on the 2D interaction map. 

\subsubsection{Training Details} We optimize ConvNCF with the BPR objective with mini-batch Adagrad~\cite{duchi2011adaptive}. Specifically, in each epoch, we first shuffle all observed interactions, and then get a mini-batch in a sequential way;
given the mini-batch of observed interactions, we then generate negative examples on the fly to get the training triplets. The negative examples are randomly sampled from a uniform distribution; while recent efforts show that a better negative sampler can further improve the performance~\cite{www2018improvedBPR}, we leave this exploration as future work. 
We pre-train the embedding layer with MF. After pre-training, considering that other parameters of ConvNCF are randomly initialized and the overall model is in a underfitting state, we train ConvNCF for $1$ epoch first without any regularization. For the following epochs, we enforce regularization on ConvNCF, including $L_2$ regularization on the embedding layer, convolution layers, and the output layer, respectively. Note that the regularization coefficients (especially for the output layer) have a very large impact on model performance. 


\begin{table*}
\centering
\resizebox{0.93\textwidth}{!}{%
	\begin{tabular}{|l|l|l|l|l|l|l||l|l|l|l|l|l|c|}
		\hline
		& \multicolumn{6}{c||}{\textbf{Gowalla}} & \multicolumn{6}{c|}{\textbf{Yelp}} & \\\hline
		& \multicolumn{3}{c|}{\textbf{HR@$k$}} & \multicolumn{3}{c||}{\textbf{NDCG@$k$}} & \multicolumn{3}{c|}{\textbf{HR@$k$}} & \multicolumn{3}{c|}{\textbf{NDCG@$k$}} & RI\\\hline
		& $k=5$    & $k=10$   & $k=20$   & $k=5$    & $k=10$   & $k=20$   & $k=5$    & $k=10$   & $k=20$   & $k=5$    & $k=10$   & $k=20$   & \\\hline\hline
		ItemPop & 0.2003 & 0.2785 & 0.3739 & 0.1099 & 0.1350 & 0.1591 & 0.0710 & 0.1147 & 0.1732 & 0.0365 & 0.0505 & 0.0652 & +227.6\%\\\hline
MF-BPR & 0.6284 & 0.7480 & 0.8422 & 0.4825 & 0.5214 & 0.5454 & 0.1752 & 0.2817 & 0.4203 & 0.1104 & 0.1447 & 0.1796 & +9.5\%\\\hline
MLP & 0.6359 & 0.7590 & 0.8535 & 0.4802 & 0.5202 & 0.5443 & 0.1766 & 0.2831 & 0.4203 & 0.1103 & 0.1446 & 0.1792 & +9.2\%\\\hline
JRL & 0.6685 & 0.7747 & 0.8561 & 0.5270 & 0.5615 & 0.5821 & 0.1858 & 0.2922 & 0.4343 & 0.1177 & 0.1519 & 0.1877 & +3.9\%\\\hline
NeuMF & 0.6744 & 0.7793 & 0.8602 & 0.5319 & 0.5660 & 0.5865 & 0.1881 & 0.2958 & 0.4385 & 0.1189 & 0.1536 & 0.1895 & +3.0\%\\\hline
ConvNCF & \textbf{0.6914$^*$} & \textbf{0.7936$^*$} & \textbf{0.8695$^*$} & \textbf{0.5494$^*$} & \textbf{0.5826$^*$} & \textbf{0.6019$^*$} & \textbf{0.1978$^*$} & \textbf{0.3086$^*$} & \textbf{0.4430$^*$} & \textbf{0.1243$^*$} & \textbf{0.1600$^*$} & \textbf{0.1939$^*$} & -\\\hline
	\end{tabular}}
\caption{Top-$k$ recommendation performance where $k\in\{5,10,20\}$. RI indicates the average improvement of ConvNCF over the baseline. $^*$ indicates that the improvements over all other methods are statistically significant for $p<0.05$.}
\label{tab:performance}
\end{table*}

\section{Experiments}
To comprehensively evaluate our proposed method, we conduct experiments to answer the following research questions:
\begin{description}
	\item[RQ1]~Can our proposed ConvNCF outperform the state-of-the-art recommendation methods?
	\item[RQ2]~Are the proposed outer product operation and the CNN layer helpful for learning from user-item interaction data and improving the recommendation performance?
	\item[RQ3]~How do the key hyperparameter in CNN (i.e., number of feature maps) affect ConvNCF's performance?
\end{description}

\subsection{Experimental Settings}

\paragraph{Data Descriptions.}
We conduct experiments on two publicly accessible datasets: Yelp\footnote{https://github.com/hexiangnan/sigir16-eals} and Gowalla\footnote{http://dawenl.github.io/data/gowalla\_pro.zip}.

\textbf{Yelp}.~This is the Yelp Challenge data for user ratings on businesses.
We filter the dataset following by~\cite{he2016fast}. Moreover, we merge the repetitive ratings at different timestamps to the earliest one, so as to study the performance of recommending novel items to a user.
The final dataset obtains 25,815 users, 25,677 items, and 730,791 ratings.

\textbf{Gowalla}.~This is the check-in dataset from Gowalla, a location-based social network, constructed by~\cite{liang2016modeling} for item recommendation.
To ensure the quality of the dataset, we perform a modest filtering on the data, retaining users with at least two interactions and items with at least ten interactions.
The final dataset contains 54,156 users, 52,400 items, and 1,249,703 interactions. \vspace{+5pt}

\paragraph{Evaluation Protocols.}
For each user in the dataset, we holdout the latest one interaction as the testing positive sample, and then pair it with $999$ items that the user did not rate before as the negative samples.
Each method then generates predictions for these $1,000$ user-item interactions.
To evaluate the results, we adopt two metrics \textit{Hit Ratio} (HR) and \textit{Normalized Discounted Cumulative Gain} (NDCG), same as  \cite{he2017neural}.
HR@$k$ is a recall-based metric, measuring whether the testing item is in the top-$k$ position (1 for yes and 0 otherwise).
NDCG@$k$ assigns the higher scores to the items within the top $k$ positions of the ranking list.
To eliminate the effect of random oscillation, we report the average scores of the last ten epochs after convergence. \vspace{+5pt}

\paragraph{Baselines.}
To justify the effectiveness of our proposed ConvNCF, we study the performance of the following methods:

\textbf{1. ItemPop} ranks the items based on their popularity, which is calculated by the number of interactions. It is always taken as a benchmark for recommender algorithms.

\textbf{2. MF-BPR}~\cite{rendle2009bpr} optimizes the standard MF model with the pairwise BPR ranking loss.

\textbf{3. MLP}~\cite{he2017neural} is a NCF method that concatenates user embedding and item embedding to feed to the standard MLP for learning the interaction function. 

\textbf{4. JRL}~\cite{zhang2017joint} is a NCF method that places a MLP above the element-wise product of user embedding and item embedding. Its difference with GMF~\cite{he2017neural} is that JRL uses multiple hidden layers above the element-wise product, while GMF directly outputs the prediction score. 

\textbf{5. NeuMF}~\cite{he2017neural} is the state-of-the-art method for item recommendation, which combines hidden layer of GMF and MLP to learn the user-item interaction function. \vspace{+5pt}

\paragraph{Parameter Settings.}
We implement our methods with Tensorflow, which is available at: \href{https://github.com/duxy-me/ConvNCF}{https://github.com/duxy-me/ConvNCF}. We randomly holdout 1 training interaction for each user as the validation set to tune hyperparameters.
We evaluate ConvNCF of the specific setting as illustrated in Figure~\ref{fig:ConNCF}.
The regularization coefficients are separately tuned for the embedding layer, convolution layers, and output layer in the range of $[10^{-3}, 10^{-2}, ..., 10^2]$.  
For a fair comparison, we set the embedding size as 64 for all models and optimize them with the same BPR loss using mini-batch Adagrad (the learning rate is 0.05).
For MLP, JRL and NeuMF that have multiple fully connected layers, we tuned the number of layers from 1 to 3 following the tower structure of \cite{he2017neural}. 
For all models besides MF-BPR, we pre-train their embedding layers using the MF-BPR, and the $L_2$ regularization for each method has been fairly tuned. 

\subsection{Performance Comparison (RQ1)}

	Table~\ref{tab:performance} shows the Top-$k$ recommendation performance on both datasets where $k$ is set to 5, 10, and 20.
	We have the following key observations:
	\begin{itemize}[leftmargin=*]
	    \item ConvNCF achieves the best performance in general, and obtains high improvements over the state-of-the-art methods. This justifies the utility of ONCF framework that uses outer product to obtain the 2D interaction map, and the efficacy of CNN in learning high-order correlations among embedding dimensions. 
		\item JRL consistently outperforms MLP by a large margin on both datasets.
		This indicates that, explicitly modeling the correlations of embedding dimensions is rather helpful for the learning of the following hidden layers, even for simple correlations that assume dimensions are independent of each other. Meanwhile, it reveals the practical difficulties to train MLP well, although it has strong representation ability in principle~\cite{hornik1991approximation}.
	\end{itemize}
	
	\subsection{Efficacy of Outer Product and CNN (RQ2)}
	Due to space limitation, for the blow two studies, we only show the results of NDCG, and the results of HR admit the same trend thus they are omitted. 
	
\begin{figure}[t!]
		\centering
		\includegraphics[width=\linewidth]{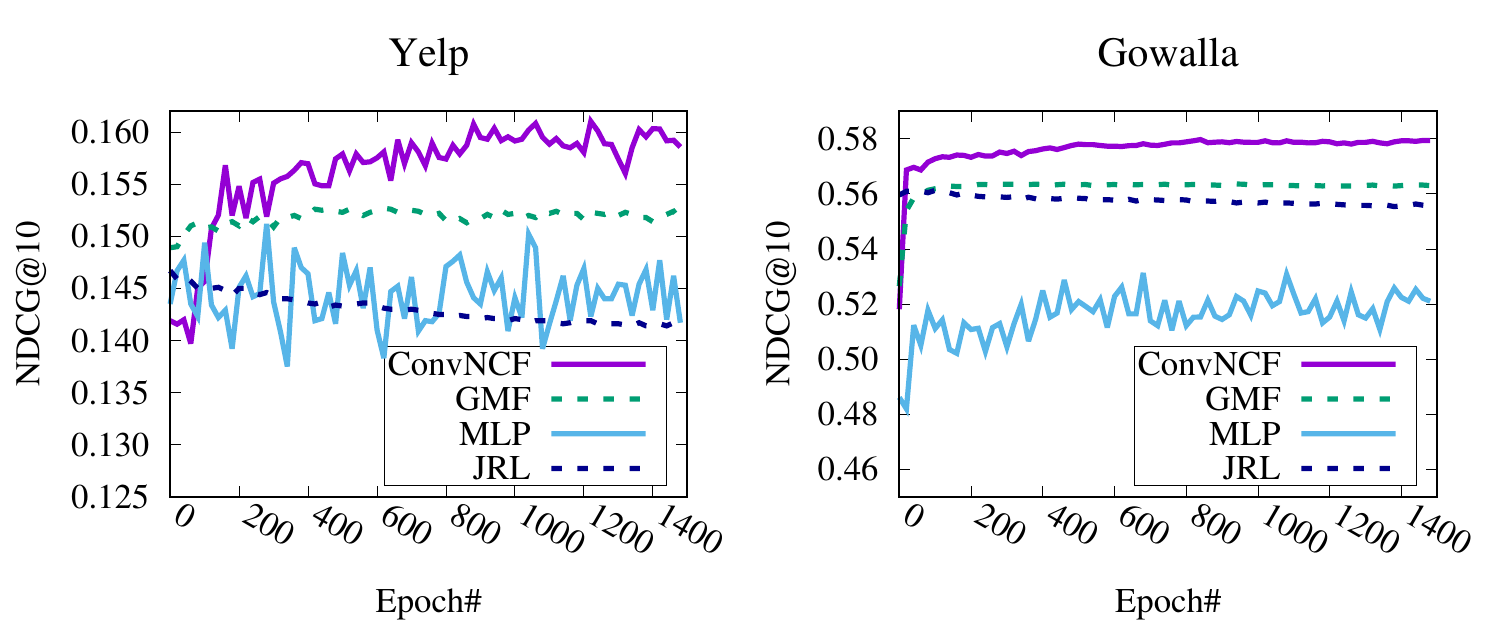}
		\caption{NDCG@10 of applying different operations above the embedding layer in each epoch (GMF and JRL use element-wise product, MLP uses concatenation, and ConvNCF uses outer product). }\vspace{-10px}
		\label{fig:ovi}
	\end{figure}

	\paragraph{Efficacy of Outer Product.}
	To show the effect of outer product, we replace it with the two common choices in existing solutions --- concatenation (i.e., MLP) and element-wise product~(i.e., GMF and JRL). We compare their performance with ConvNCF in each epoch in Figure \ref{fig:ovi}. We observe that ConvNCF outperforms other methods by a large margin on both datasets, verifying the positive effect of using outer product above the embedding layer. Specifically, the improvements over GMF and JRL demonstrate that explicitly modeling the correlations between different embedding dimensions are useful. Lastly, the rather weak and unstable performance of MLP imply the difficulties to train MLP well, especially when the low-level has fewer semantics about the feature interactions. This is consistent with the recent finding of \cite{NFM} in using MLP for sparse data prediction. \vspace{+5pt}. 
	
	\begin{figure}[t!]
		\centering
		\includegraphics[width=\linewidth]{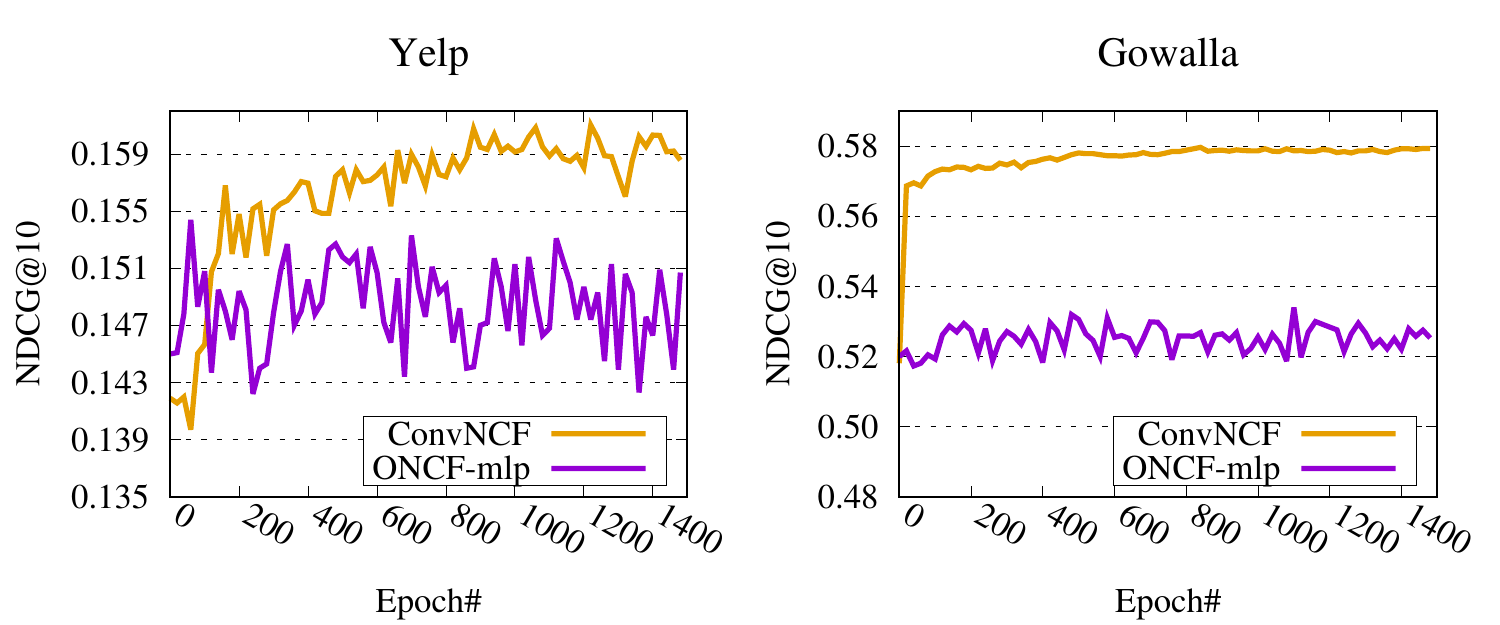}
		\caption{NDCG@10 of using different hidden layers for ONCF (ConvNCF uses a 6-layer CNN and ONCF-mlp uses a 3-layer MLP above the interaction map). }\vspace{-10px}
		\label{fig:cnn}
	\end{figure}

	\paragraph{Efficacy of CNN.} To make a fair comparison between CNN and MLP under our ONCF framework, we use MLP to learn from the same interaction map generated by outer product. Specifically, we first flatten the interaction as a $K^2$ dimensional vector, and then place a 3-layer MLP above it. We term this method as ONCF-mlp. Figure~\ref{fig:cnn} compares its performance with ConvNCF in each epoch. We can see that ONCF-mlp performs much worse than ConvNCF, in spite of the fact that it uses much more parameters (3 magnitudes) than ConvNCF. Another drawback of using such many parameters in ONCF-mlp is that it makes the model rather unstable, which is evidenced by its large variance in epoch. In contrast, our ConvNCF achieves much better and stable performance by using the locally connected CNN. These empirical evidence provide support for our motivation of designing ConvNCF and our discussion of MLP's drawbacks in Section~\ref{ssec:cncf}.
	
	
	\subsection{Hyperparameter Study (RQ3)}
	
\begin{figure}[t!]
	\centering
	\includegraphics[width=\linewidth]{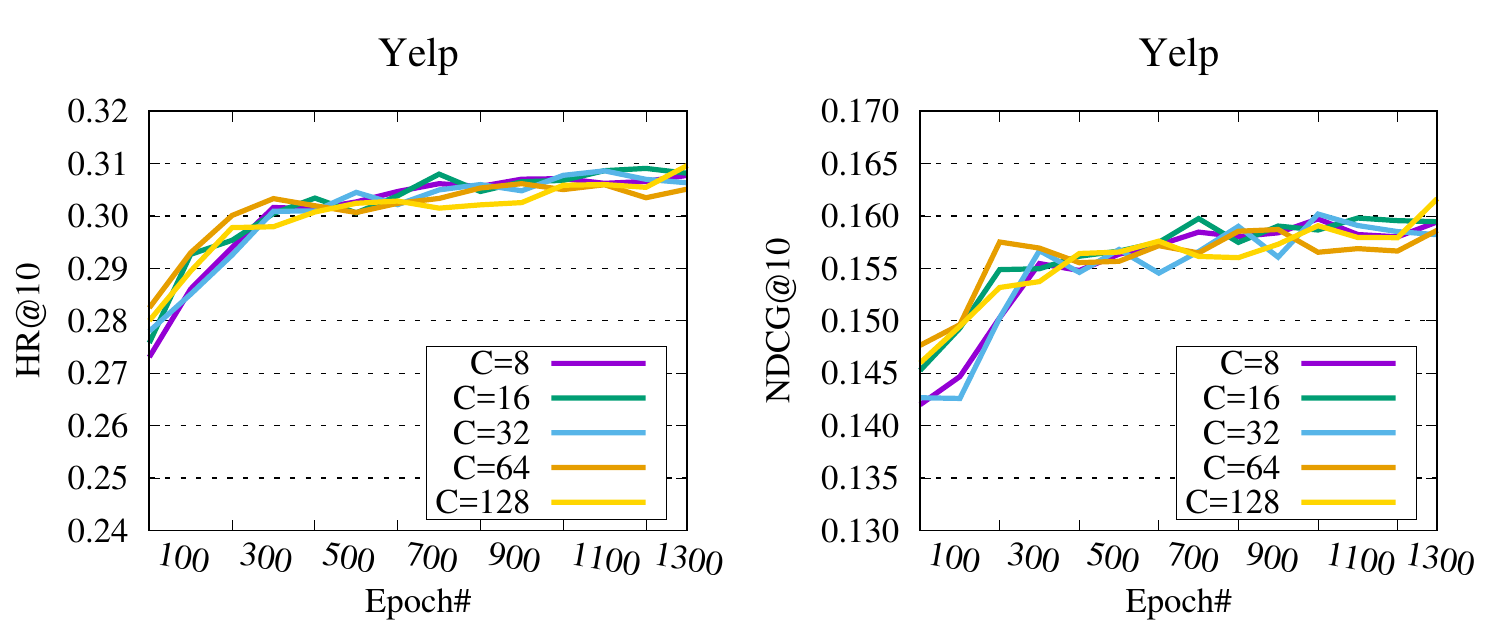}
	\vspace{-10px}
	\caption{Performance of ConvNCF w.r.t. different numbers of feature maps per convolutional layer (denoted by $C$) in each epoch on Yelp. }\vspace{-10px}
	\label{fig:layer}
\end{figure}

	\paragraph{Impact of Feature Map Number.} The number of feature maps in each CNN layer affects the representation ability of our ConvNCF. 
	Figure \ref{fig:layer} shows the performance of ConvNCF with respect to different numbers of feature maps.
	We can see that all the curves increase steadily and finally achieve similar performance, though there are some slight differences on the convergence curve. This reflects the strong expressiveness and generalization of using CNN under the ONCF framework since dramatically increasing the number of parameters of a neural network does not lead to overfitting. Consequently, our model is very suitable for practical use.



\section{Conclusion}
We presented a new neural network framework for collaborative filtering, named ONCF. 
The special design of ONCF is the use of an outer product operation above the embedding layer, which results in a semantic-rich interaction map that encodes pairwise correlations between embedding dimensions. This facilitates the following deep layers learning high-order correlations among embedding dimensions. To demonstrate this utility, we proposed a new model under the ONCF framework, named ConvNCF, which uses multiple convolution layers above the interaction map. Extensive experiments on two real-world datasets show that ConvNCF outperforms state-of-the-art methods in top-$k$ recommendation. In future, we will explore more advanced CNN models such as ResNet~\cite{he2016deep} and DenseNet~\cite{huang2017densely} to further explore the potentials of our ONCF framework. Moreover, we will extend ONCF to content-based recommendation scenarios~\cite{ACF,Yu:2018:ACR}, where the item features have richer semantics than just an ID. Particularly, we are interested in building recommender systems for multimedia items like images and videos, and textual items like news. \vspace{+5pt}

\section{Acknowledgments}
This work is supported by the National Research Foundation, Prime Minister's Office, Singapore under its IRC@SG Funding Initiative, 
by the 973 Program of China under Project No.: 2014CB347600, by the Natural Science Foundation of China under Grant No.: 61732007, 61702300, 61501063, 61502094, and 61501064, by the Scientific Research Foundation of Science and Technology Department of Sichuan Province under Grant No. 2016JY0240, and by the Natural Science Foundation of Heilongjiang Province of China (No.F2016002). Jinhui Tang is the corresponding author. 

\appendix

\bibliographystyle{named}
\bibliography{ijcai18}

\end{document}